# Computational prediction of ferromagnetic $AT_6X_6$ kagome compounds

Shiya Chen[1], Zhen Zhang[2], Vladimir Antropov[2,3†], and Yang Sun[1*]

[1]Department of Physics, Xiamen University, Xiamen 361005, China
[2]Department of Physics and Astronomy, Iowa State University, Ames, IA 50011, USA
[3]Ames National Laboratory, Ames, Iowa 50011, USA

We present a systematic high-throughput density-functional-theory investigation of the structural and magnetic stability of 312 substitutional compounds in the magnetic kagome $AT_6X_6$ family. Our screening confirms the stability of many previously reported structures and predicts several additional stable candidates. Within collinear spin configurations, we find that Fe-based systems predominantly adopt antiferromagnetic ground states, whereas Mn-based analogues exhibit a more balanced distribution between ferromagnetic and antiferromagnetic order. For compounds exhibiting several nearly degenerate collinear configurations, we analyze the nature of their magnetic ground states, assess the possible emergence of non-collinear order, and discuss the limitations and uncertainties inherent to standard density-functional approaches. Our electronic-structure analysis further reveals that newly predicted ferromagnetic kagome systems display characteristic features of topological metals, with rich magnetic configurations that can be tuned by chemical substitution. Overall, these ferromagnetic kagome compounds constitute a broad and still largely unexplored materials platform for the emergence of exciting magneto-transport phenomena.

## I. Introduction

The kagome lattice, a two-dimensional network of corner-sharing triangles, has emerged as fertile ground for realizing magnetic topological quantum states [1,2]. Its unique geometry naturally gives rise to a signature electronic band structure featuring Dirac fermions, correlation-hosting flat bands (FB), and van Hove singularities (VHS) [3]. The combination of these features with magnetic order and spin-orbit coupling (SOC) can generate substantial Berry curvature, driving a range of exotic quantum transport phenomena, including large anomalous Hall (AHE) and Nernst effects [4,5].

Among kagome platforms, the $AT_6X_6$ (where A denotes the spacer cation, T the transition metal on the kagome sites, and X a p-block element) family is especially appealing. Members of this family host quasi-2D kagome layers of T atoms that are effectively isolated by A/X spacer blocks, helping preserve intrinsic kagome band features. While the substitutions on A, T, and X sites offer vast chemical tunability [6,7], they also foster a complex hierarchy of magnetic interactions [8, 9]: Fe-based systems favor antiferromagnetism [10], whereas Mn-based analogues exhibit complex noncollinear (NC) spiral states [8, 10], frustrating efforts to realize a clean FM ground state. While the profound potential of such FM systems was clearly demonstrated by the few known examples, such as $TbMn_6Sn_6$ [11, 12] and $LiMn_6Sn_6$ [13–15], which exhibit large AHE and Weyl-like band features, the discovery of new FM $AT_6X_6$ has been sporadic rather than systematic. A comprehensive high-throughput (HTP) study to navigate this complex magnetic phase space has been conspicuously absent.

In parallel, HTP first-principles screening has already emerged as a powerful route to discovering functional quantum materials in many contexts, including conventional boride superconductors, magnetic borides, and altermagnets, particularly when tightly coupled to targeted synthesis and characterization [16–23]. Building on these successes, we now bring the HTP paradigm to kagome $AT_6X_6$ intermetallics, aiming to systematically map their structural and magnetic phase space and to pinpoint promising ferromagnetic platforms for experimental exploration.

In this work, we carry out a systematic HTP investigation of the structural and magnetic stability of a broad set of elemental substitutions across several key $AT_6X_6$ structural prototypes. This survey not only reproduces known trends but also reveals several previously unreported stable compounds, thereby providing a practical roadmap for experimental validation. Focusing on a subset of representative predictions, we examine the applicability of a Heisenberg model , establish a quantitative criterion for assessing the reliability of our magnetic ground-state predictions, and demonstrate how their electronic structures can be leveraged to identify characteristic topological band features.

## II. Methods

The substituted $AT_6X_6$ structures were optimized by spin-polarized density functional theory (DFT) calculations using the VASP code [24,25], which employs the projector augmented wave (PAW) method [26]. The exchange and correlation energy is treated with the spin-polarized generalized gradient approximation (GGA) and parameterized by the Perdew−Burke−Ernzerhof formula (PBE) [27]. A plane-wave basis was used with a kinetic energy cutoff of 520 eV, and the convergence criterion for the total energy was set to $10^{-5}$ eV. Monkhorst−Pack's sampling scheme [28] was adopted for Brillouin-zone sampling with a k-point spacing of $2\pi \times 0.033$ Å$^{-1}$. The lattice and atomic coordinates are fully relaxed until the force on each atom is less than 0.01 eV/Å. All geometry optimizations were performed in the FM configuration. The energies of the different magnetic states were obtained from self-consistent calculations performed directly on the FM-relaxed structures. We additionally performed AFM relaxations for two representative stable compounds and found that the structural changes from FM to AFM are minor (Table S5). SOC effect was included only in the post-processing of electronic structures.

†Email: antropov@ameslab.gov
*Email: yangsun@xmu.edu.cn

The formation energy $E_f$ of each ternary $AT_6X_6$ was calculated by

$$E_f(\mathrm{AT_6X_6}) = \frac{E(\mathrm{AT_6X_6}) - E(\mathrm{A}) - 6E(\mathrm{T}) - 6E(\mathrm{X})}{13} \quad (1)$$

where $E(\mathrm{AT_6X_6})$ is the energy of $AT_6X_6$. $E$(A), $E$(T), and $E$(X) are the energy of A, T, and X ground-state bulk phases, respectively. We characterized the thermodynamic stability of each $AT_6X_6$ compound by calculating its energy above the convex hull $E_{\mathrm{d}}$, which represents the energy difference against the most stable competing reference phases. The competing reference phases for the convex-hull construction were identified from the Materials Project database [29]. Their total energies were then recalculated using the same computational setup employed in this work, ensuring an internally consistent dataset for the formation-energy and convex-hull analysis. The Heisenberg model stability parameters have been obtained using the code from Ref. [30,31], with the OPENMX package.

Phonon spectrum was computed using the finite-displacement method implemented in the PHONOPY code [32] The force constants were computed from DFT calculations on the fully relaxed magnetic ground-state structures. For each compound, a $2 \times 2 \times 2$ supercell containing 104 atoms was employed and a displacement amplitude of 0.01 Å was applied. The elastic constants were calculated using the stress-strain relation, where small symmetry-preserving strains were applied to the fully relaxed structures and the resulting stress responses were used to obtain the elastic tensor.

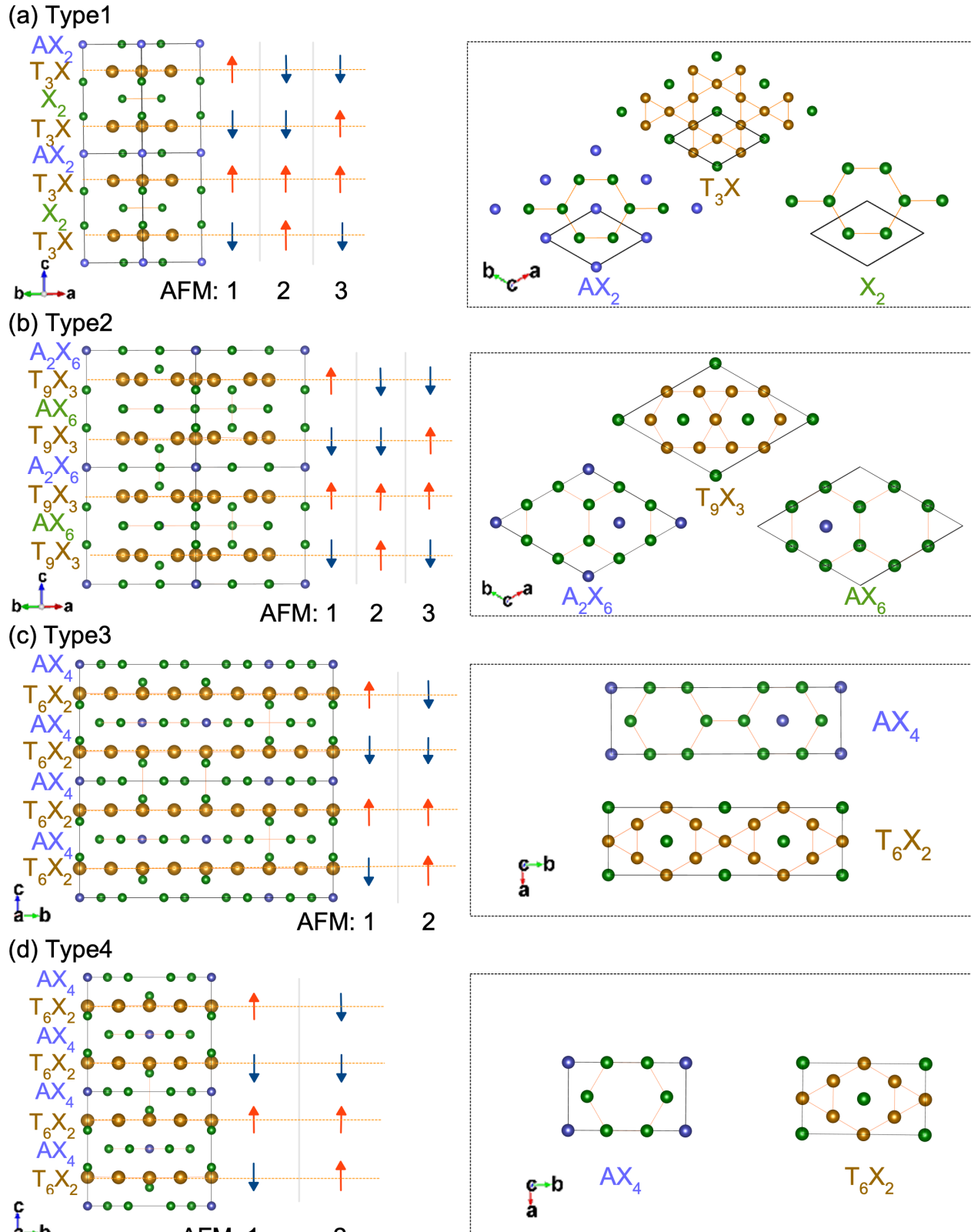

**FIG. 1** Schematic illustrations of the four structural prototypes and the collinear magnetic configurations considered in this work. Panels (a)–(d) correspond to Type 1–Type 4, respectively. In the left part of each panel, the crystal stacking sequence along the out-of-plane direction is shown, where A, T, and X atoms are represented by blue, orange, and green spheres, respectively. The orange dashed lines mark the magnetic T-containing layers, and the red/blue arrows indicate the spin orientations used to define the different antiferromagnetic configurations (AFM1–AFM3 for Types 1 and 2; AFM1–AFM2 for Types 3 and 4). The right part of each panel shows the corresponding projected atomic arrangements of the constituent structural blocks, including $AX_2$, $T_3X$, and $X_2$ for Type 1; $AX_2$, $T_9X_3$, and $AX_6$ for Type 2; and $AX_4$ and $T_6X_2$ for Types 3 and 4. The viewing directions are indicated by the coordinate axes in each panof the corresponding structural motifs. Purple, gold, and green spheres denote A, T, and X atoms, respectively.

## III. Results and discussions

### A. Stability screening for the AT6X6 system

We adopt four crystal structures as templates for the $AT_6X_6$ family and perform systematic elemental substitutions. These structures share a common layered framework: transition-metal (T) kagome monolayers are separated by A-atom and X-atom honeycomb spacers. As shown in Fig. 1, the first two configurations in Fig. 1(a) and (b) are hexagonal polymorphs with *P6/mmm* symmetry. Type 1, representing the experimental $HfFe_6Ge_6$ structure, consists of fully A-filled $AX_2$ layers alternating with vacant $X_2$ honeycomb layers along the c-axis. Type 2, corresponding to the experimental $LiFe_6Ge_6$ structure, is an in-plane $\sqrt{3} \times \sqrt{3}$ superstructure of the type 1; its key feature is that the A-atoms exhibit different levels of partial occupancy in the alternating $A_2X_6$ and $AX_6$ spacer layers.

The remaining two configurations in Fig. 1(c) and (d) are orthorhombic. In both type 3 (*Cmcm*, e.g., the experimental $TbFe_6Sn_6$ system) and type 4 (*Immm*, e.g., the experimental $ScFe_6Ga_6$ system), A-atoms occupy every spacer layer, maintaining an identical per-layer occupancy. The primary distinction lies in their in-plane ordering within the honeycomb centers: in the type 3 structure, the A-atoms occupy the sites in a 'double-column' (two-by-two) arrangement, whereas in the type 4, they occupy the sites in an alternating 'single-column' (one-by-one) pattern.

For the $AT_6X_6$ compositional screening, we restrict the transition-metal site T to Fe and Mn. For the X site, Ge and Sn are well-established experimentally, and neighboring p-block elements (e.g., Ga) were explored as ±1-electron analogs. The A-site selection began with known stable members (Li, Mg, Sc, Y, Nb, Zr) and was chemically extended to adjacent elements (e.g., Na, K, Rb, Sr, La, Ti, Hf). This combinatorial design yields 312 $AT_6X_6$ configurations in total. The stability data for all systems is summarized in Fig. 2 (we also tested X = In, Pb, As, Sb, but these variants are generally unstable with positive formation energies; see Fig S1). To prioritize compounds with synthesis potential, we classify materials with $E_d < 100$ meV/atom as metastable, and those with $E_d < 5$ meV/atom as stable, which heuristically accounts for finite-temperature effects and functional-related DFT errors. Compounds satisfying this criterion are considered promising synthesis candidates. Comparable convex-hull tolerances have been adopted in prior HTP studies [33,34]. Applying this filter yields 233 metastable phases. A full list of these metastable phases, their structural prototypes, and $E_\mathrm{d}$ values is provided in Table S6.

As shown in Fig. 2, compounds with X = Ge or Sn preferentially adopt the type 1 structure. This set includes the experimentally synthesized Fe-based $AFe_6Ge_6$ (A = Mg, Sc, Y, Ti, Zr, Hf, Nb) and $AFe_6Sn_6$ (A = Sc, Y, Zr) compounds, along with their Mn-based analogs $AMn_6Ge_6$ (A = Mg, Sc, Y, Zr, Hf) and $AMn_6Sn_6$ (A = Li, Mg, Sc, Zr, Y, Hf). Although experimental formation-enthalpy data are very limited for this family of intermetallic kagome compounds, our calculations therefore reproduce the experimentally reported representatives as stable or metastable, in agreement with prior crystallographic studies. Notably, our screening also uncovers several stable phases that have not yet been experimentally observed, namely $NaFe_6Ge_6$, $LiMn_6Ge_6$, $NaMn_6Ge_6$, $LaMn_6Ge_6$, $TiMn_6Ge_6$ and $NbMn_6Ge_6$. The corresponding convex-hull phase diagrams and competing reference phases for these newly predicted stable compounds are provided in Supplementary Fig. S4 and Supplementary Table S7. However, we note a significant discrepancy for $LiFe_6Ge_6$: the experimentally synthesized phase is reported as type 2 [35], which differs from our calculated ground state, the type 4 structure.

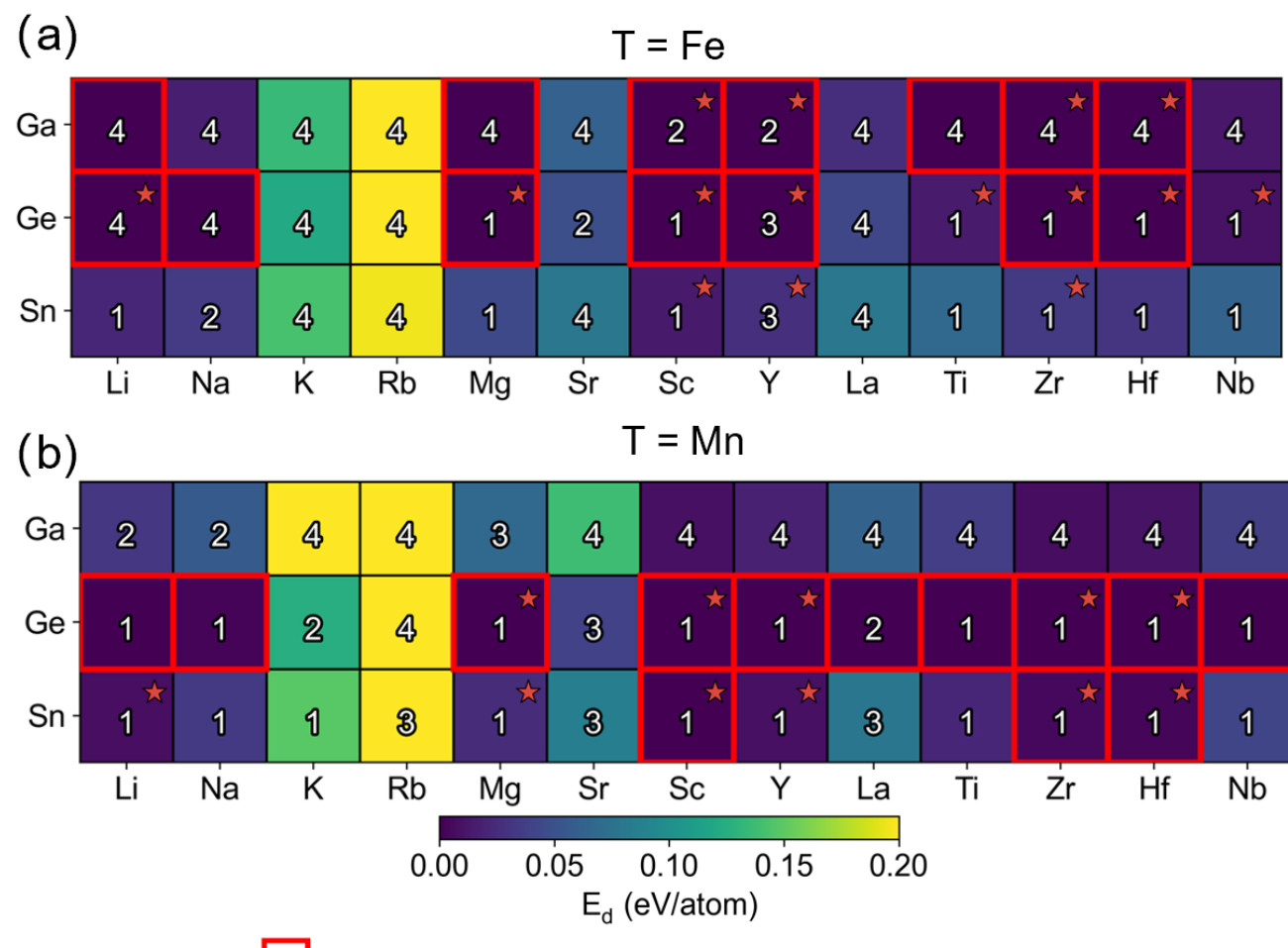

**FIG. 2** The $E_d$ for (a) $AFe_6X_6$ and (b) $AMn_6X_6$. The horizontal axis lists A-site elements from Li to Nb; the vertical axis lists X-site elements in the order Ga, Ge, and Sn. The number in each grid denotes the most stable structural motif—1 (type 1 structure), 2 (type 2 structure), 3 (type 3 structure) or 4 (type 4 structure)—identified for that composition. Red borders indicate $E_\mathrm{d} < 5$ meV/atom (on the convex hull) and the red stars indicate that system with this composition has already been experimentally synthesized.

In contrast, for X = Ga, our calculations show that most compositions favor the type 4 structure. This finding is consistent with previous experimental reports, as the known $AT_6Ga_6$ systems—including $ScFe_6Ga_6$, $YFe_6Ga_6$, $ZrFe_6Ga_6$ and $HfFe_6Ga_6$—all adopt the type 4 structure. Furthermore, our screening identifies three new systems, $LiFe_6Ga_6$, $MgFe_6Ga_6$ and $TiFe_6Ga_6$, which are also predicted to be stable in the type 4 structure. However, a similar situation to $LiFe_6Ge_6$ case arises here: for both $ScFe_6Ga_6$ and $YFe_6Ga_6$, our calculations find that the type 2 is energetically more favorable than the observed type 4.

Although our predicted ground-state structures for $ScFe_6Ga_6$, $YFe_6Ga_6$, and $LiFe_6Ge_6$ do not match the experimentally reported phases, the energy differences between them are extremely small, at only 0.1, 0.009, and 1.3 meV/atom, respectively. Such minute energy gaps fall within the numerical and physical uncertainty typical of layered polytypes and can be easily reversed by accounting for finite-temperature vibrational or magnetic entropy. In addition, changes in exchange-correlation functionals or interlayer dispersion may also reshuffle the energetic ordering. We thus infer that $LiFe_6Ge_6$, $ScFe_6Ga_6$, and $YFe_6Ga_6$ host near-degenerate stacking polytypes, where synthesis conditions, annealing, slight off-stoichiometry, or stacking faults likely stabilize the observed forms.

This coexistence of multiple polymorphs is strongly supported by the $YFe_6Sn_6$ system. A complete summary of the energy difference between the lowest-energy and second-lowest-energy structural configurations is provided in Fig. S2. For this compound, our predicted ground state is the type 3, which has indeed been synthesized [36]. However, the type 4 is only 1 meV/atom higher in energy, and has also been experimentally realized [37], differing only in its synthesis method. Beyond $YFe_6Sn_6$, our results indicate that this structural flexibility is a pervasive feature of the $AT_6X_6$ family. As detailed in Table S6, we identify multiple competitive stable phases for several other systems—such as $YFe_6Ge_6$ (Types 1, 3, and 4), $NaFe_6Ge_6$ (Types 1 and 4), and $MgMn_6Ge_6$ (Types 1 and 3)—all of which exhibit energy differences well within the 5 meV/atom stability window. Therefore, it is highly probable that the ground states we predicted, even those not yet matching experimental reports, are physically plausible and could be synthesized. This behavior also aligns with prior observations of intergrowth and long-period ordering in the $ScFe_6Ge_{6-x}Ga_x$ series, where closely related stacking variants coexist and continuously repartition with composition—classic hallmarks of meV-per-atom near-degeneracy [38]. These structural trends provide the framework for analyzing their magnetic ground states in the next section.

## B. Magnetic Ground States

Experimentally, $AT_6X_6$ systems have been extensively studied, and a variety of magnetic orders have been determined. In addition to the usual FM and AFM orderings, various complex spin-spiral (helical) magnetic structures have been established by neutron-scattering experiments. Thus, determining and explaining the magnetic ground state can be very challenging , because to properly identify the nature of such order one should analyze effects such as frustrated isotropic Heisenberg exchange, non-Heisenberg biquadratic and higher-order exchange terms, anisotropic exchange, magnetic anisotropy, and magnetostriction. A full treatment of these effects is beyond the scope of a HTP survey. Moreover, performing NC or spin-spiral calculations for more than 300 compounds remains computationally prohibitive.

Nevertheless, collinear calculations provide a practical and well-established starting point: they allow us to map the dominant magnetic tendencies, identify frustrated cases, and flag systems that are susceptible to noncollinearity. We therefore restrict the screening to collinear configurations and establish simple criteria under which NC instabilities are expected to emerge. To determine the collinear magnetic ground state, we compared the energies of a FM state against several AFM configurations. These configurations, denoted AFM1, AFM2, and AFM3, are defined by their interlayer stacking sequence along the c-axis. All three retain FM alignment within the kagome planes, as defined schematically in the middle panels of Fig. 1. The predicted collinear magnetic ground states for all substituted systems are visualized in the heatmap in Fig. 3. A detailed breakdown of the magnetic ground state, and the energy difference between FM and the most stable AFM phase ($\Delta E$) for all experimentally known and newly predicted stable compounds is provided in Table 1. Hereafter, the unit 'meV' refers to meV per transition metal atom when describing magnetic energy differences.

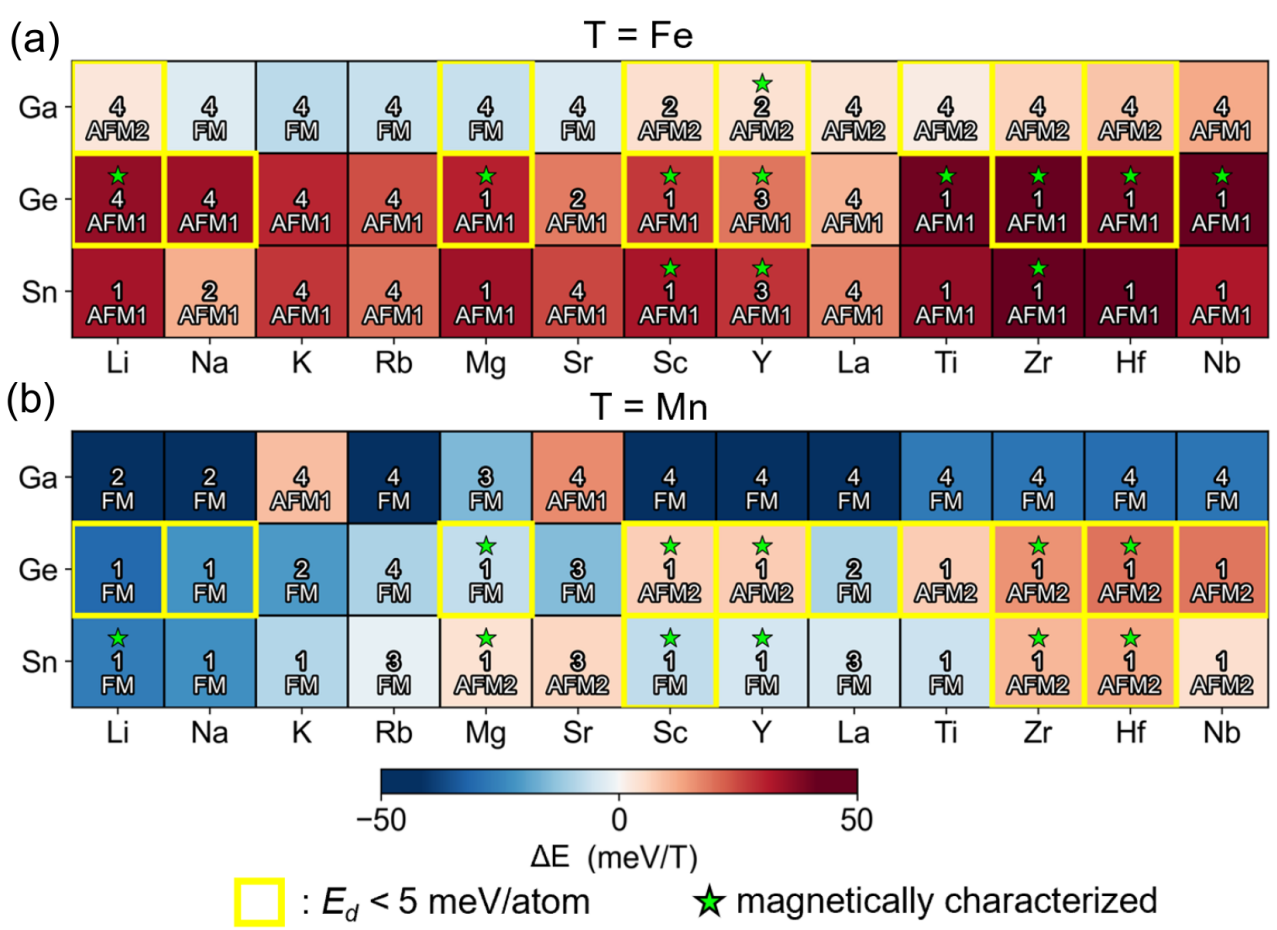

**FIG. 3** (a-b) Calculated magnetic ground states of $AT_6X_6$ compounds (similar to Fig.2). The upper label marks the most stable structural template. The lower label indicates the collinear magnetic ground state from DFT screening. Color encodes the energy difference between the FM and the lowest-energy AFM states, $\Delta E = E(FM) - E_{min}(AFM)$, in meV (negative favors FM; positive favors AFM). Unless otherwise noted, all magnetic energy differences are given in meV per transition metal atom. Stars denote the system where magnetism was previously identified.

For X = Ge and Sn, Fe-based $AT_6X_6$ compounds overwhelmingly favor an AFM1 ground state as shown in Fig. 3(a). This trend is consistent with numerous previous experimental reports [7,39–42] and theoretical prediction for $LiFe_6Ge_6$ [43]. However, significant R-site vacancies (~10–20%) reported in experimentally studied $ScFe_6Sn_6$ and $ZrFe_6Sn_6$ can lead to local magnetic frustration or spin-glass-like behavior [42]. We find that the newly identified stable type 4 $NaFe_6Ge_6$ also follows this trend, with its AFM1 state being 29.4 meV lower in energy than the FM state. By contrast, for X = Ga, our calculations favor AFM2 or FM orders, whereas explicit assignments of collinear ground states for $AFe_6Ga_6$ systems remain scarce in the literature. Although several type 4 $AFe_6Ga_6$ (A = Sc, Y, Zr, Hf) compounds have been experimentally synthesized, definitive magnetic structure data is practically absent. For instance, while magnetic measurements for type 4 *Immm* $YFe_6Ga_6$ exist, these reports only broadly classify it as ferrimagnetic (FIM) without determining the specific magnetic ordering [44]. For this system, FM behavior was also reported in *I4/mmm* solid-solution phase [45,46]; however, as our present study is restricted to ordered polymorphs, that phase is not discussed further.

**Table 1** Stabilities and magnetic properties of experimentally synthesized (marked with *) and newly predicted stable ($E_d < 5$ meV/atom) $AT_6X_6$ compounds. The table lists magnetic moments and the magnetic ground state, with experimentally reported states shown in parentheses. Magnetic moments are given in $\mu_B$ per transition metal atom. The energy difference is calculated as $\Delta E = E(FM) - E_{min}(AFM)$, where $E_{min}(AFM)$ is the energy of the lowest-energy AFM configuration (AFM if $\Delta E > 0$ and FM if $\Delta E < 0$.) Systems with $|\Delta E| > 15$ meV are highlighted to indicate a robust collinear magnetic ground state.

| No. | Struct. type | System | $E_d$ (meV/atom) | Magnetic Ground State | Moment/$\mu_B$ | $\Delta E$/meV |
|---|---|---|---|---|---|---|
| 1 | 4 | Hf-Fe-Ga | 0.0* | AFM2 | 1.89 | 6.2 |
| 2 | 1 | Hf-Fe-Ge | 0.0* | AFM1(AFM1) [39,49] | 2.02 | **33.9** |
| 3 | 2 | Li-Fe-Ge | 1.6* | AFM1(AFM) [50] | 1.68 | **34.8** |
| 4 | 1 | Mg-Fe-Ge | 0.0* | AFM1(AFM1) [40] | 1.77 | **26.2** |
| 5 | 1 | Nb-Fe-Ge | 9.3* | AFM1(AFM1) [49] | 2.07 | **43.1** |
| 6 | 1 | Sc-Fe-Ge | 0.0* | AFM1(AFM1) [49] | 1.94 | **22.5** |
| 7 | 1 | Sc-Fe-Sn | 13.4* | AFM1(AFM1) [42] | 2.29 | **27.8** |
| 8 | 4 | Sc-Fe-Ga | 0.1* | AFM2 | 1.90 | 2.1 |
| 9 | 1 | Ti-Fe-Ge | 14.1* | AFM1(AFM1) [49] | 2.01 | **35.7** |
| 10 | 4 | Y-Fe-Ga | 0* | AFM2(FIM) [44,46] | 1.93 | 3.7 |
| 11 | 3 | Y-Fe-Ge | 0.0* | AFM1(AFM1) [41,51] | 2.03 | **15** |
| 12 | 4 | Y-Fe-Sn | 27.8* | AFM1(AFM1) [37] | 2.31 | **18** |
| 13 | 3 | Y-Fe-Sn | 27.7* | AFM1 | 2.31 | **23.2** |
| 14 | 4 | Zr-Fe-Ga | 0.0* | AFM2 | 1.92 | 4.4 |
| 15 | 1 | Zr-Fe-Ge | 0.0* | AFM1(AFM1) [49] | 2.06 | **37.8** |
| 16 | 1 | Zr-Fe-Sn | 32.8* | AFM1(AFM1) [42] | 2.33 | **38** |
| 17 | 1 | Mg-Mn-Ge | 0* | FM(NC) [47] | 2.08 | -4.8 |
| 18 | 1 | Mg-Mn-Sn | 24.9* | AFM2(FM) [48] | 2.49 | 2.7 |
| 19 | 1 | Li-Mn-Sn | 7.4* | FM(FM) [13] | 2.51 | **-22.2** |
| 20 | 1 | Sc-Mn-Ge | 0.0* | AFM2(NC) [52] | 1.99 | 4.7 |
| 21 | 1 | Sc-Mn-Sn | 0.0* | FM(NC) [9] | 2.44 | -5.2 |
| 22 | 1 | Y-Mn-Ge | 0.0* | AFM2(NC) [53] | 2.05 | 5.4 |
| 23 | 1 | Y-Mn-Sn | 7.8* | FM(NC) [9] | 2.49 | -3.2 |
| 24 | 1 | Zr-Mn-Ge | 0.0* | AFM2(NC) [54] | 2.04 | 12.1 |
| 25 | 1 | Zr-Mn-Sn | 3.9* | AFM2(NC) [10] | 2.50 | 7.7 |
| 26 | 1 | Hf-Mn-Ge | 0.0* | AFM2(NC) [55] | 2.02 | **15.5** |
| 27 | 1 | Hf-Mn-Sn | 2.5* | AFM2(NC) [10] | 2.48 | 9.1 |
| 28 | 2 | La-Mn-Ge | 0 | FM | 2.14 | -7.3 |
| 29 | 4 | Li-Fe-Ga | 0 | AFM2 | 1.98 | 1.6 |
| 30 | 1 | Li-Mn-Ge | 0 | FM | 2.21 | **-26.1** |
| 31 | 4 | Mg-Fe-Ga | 0 | FM | 1.95 | -4.6 |
| 32 | 4 | Na-Fe-Ge | 0 | AFM1 | 1.73 | **29.4** |
| 33 | 1 | Na-Mn-Ge | 1.9 | FM | 2.09 | **-17.7** |
| 34 | 1 | Nb-Mn-Ge | 0 | AFM2 | 2.03 | 14.9 |
| 35 | 4 | Ti-Fe-Ga | 0 | AFM2 | 1.91 | 0.9 |
| 36 | 1 | Ti-Mn-Ge | 0 | AFM2 | 2.00 | 5.4 |

Given this ambiguity, we predict the AFM2 collinear magnetic ground states for all experimentally synthesized (Y/Zr/Hf)$Fe_6Ga_6$ systems. Furthermore, we predict that the newly identified stable phases $LiFe_6Ga_6$ and $TiFe_6Ga_6$ also adopt the AFM2 state, while $MgFe_6Ga_6$ is predicted to be FM. We also find that the energy differences between the FM and the most stable AFM states in the Ga-series are consistently small, in the order of just a few meV. For Mn-based $AMn_6X_6$, many compounds are experimentally observed to host NC ground states at low temperatures, reflecting strong magnetic frustration (see Table 1, No. 17–27). For example, several Ge-based systems (e.g., Sc/Y/Zr/HfMn$_6$Ge$_6$) display collinear AFM2 order at high temperatures and transition to noncoplanar conical states on cooling. Their Sn-based analogs (Sc/Y/Zr/HfMn$_6$Sn$_6$) also become in-plane helices at low temperatures, but with distinct precursors: (Zr/Hf)Mn$_6$Sn$_6$ emerge from an AFM2 phase [10], whereas (Sc/Y)Mn$_6$Sn$_6$ remain helical over a broad temperature range [9]. Others, such as MgMn$_6$Ge$_6$, show transitions from FM to a helix [47]. Within this set, only MgMn$_6$Sn$_6$ [48] and LiMn$_6$Sn$_6$ [13] retain a simple FM ground state at low temperatures, highlighting the rarity of this desired phase. Restricting the analysis to collinear calculations, the lowest-energy solution for these experimentally NC systems typically coincide with the known high-temperature order—FM for MgMn$_6$Ge$_6$, AFM2 for (Sc/Y/Zr/Hf)Mn$_6$Ge$_6$ and (Zr/Hf)Mn$_6$Sn$_6$—whereas for YMn$_6$Sn$_6$ and ScMn$_6$Sn$_6$ the lowest-energy collinear solution is FM (despite their observed helical ground states). Furthermore, the difference in energy between FM and AFM2 states for these systems is only a few meV. This aligns with previous work of Sadhukhan et al. [56]. Within a standard GGA/LDA framework the calculation of YMn$_6$Sn$_6$ tends to favor a FM solution.

For systems with FM ground states, our results are twofold. For LiMn$_6$Sn$_6$, our calculations correctly identify the FM ground state, consistent with experiment. For MgMn$_6$Sn$_6$, however, our calculations reveal almost degenerate FM and AFM2 states, with the AFM2 configuration being lower in energy by ~2.7 meV, contradicting the experimentally observed FM state [48]. This inconsistency, along with the energetic near-degeneracy found in many observed NC systems, motivates us to address the possible formation of NC orders in more detail.

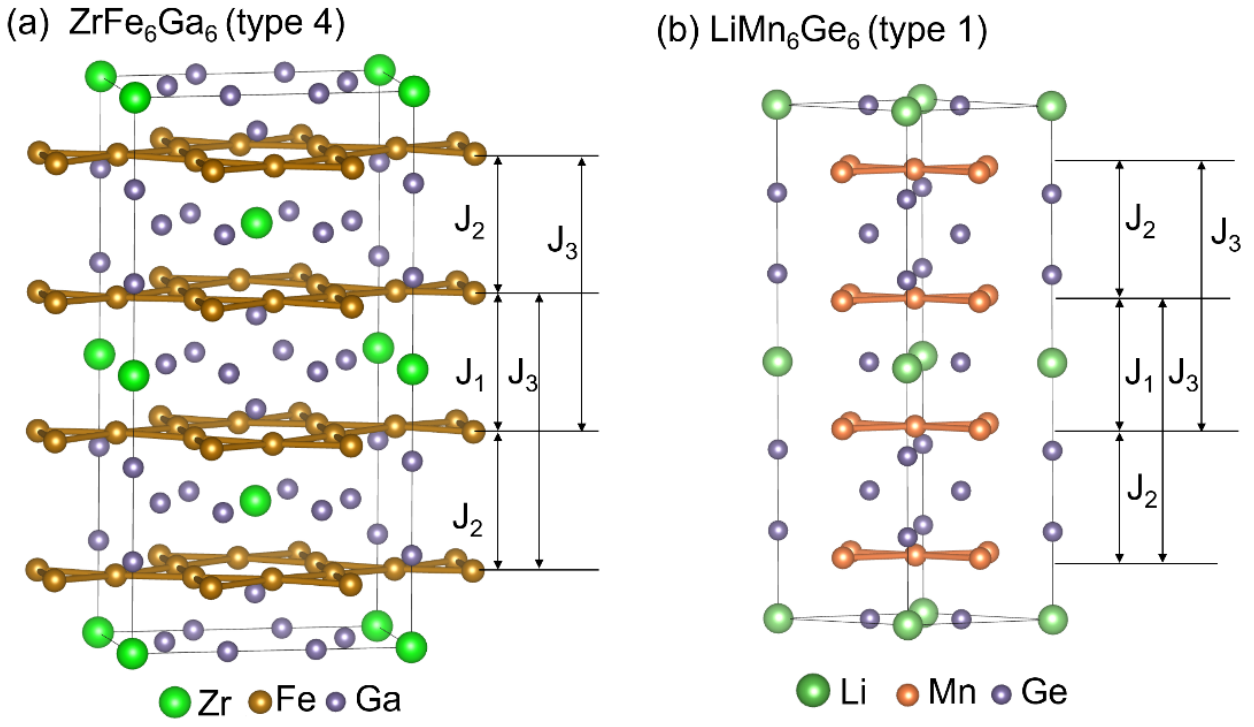

**FIG. 4** Interlayer Fe–Fe $J_n$ in type 4 ZrFe$_6$Ga$_6$ and Mn–Mn $J_n$ in type 1 LiMn$_6$Ge$_6$, $J_1$ and $J_2$ represent nearest-neighbor interlayer interactions, whereas J$_3$ represents the next-nearest-neighbor interaction. For the type 1 structure, distinct spacer layers provide the distinction between J$_1$ and J$_2$. For the type 4 structure, $J_1$ and $J_2$ distinguish parallel and antiparallel spin coupling between layers in the AFM2 configuration, respectively, with $J_1 = J_2$ for AFM1 and FM orders.

**Table 2** Values of T-T stability parameter $J_n$ for type 4 ZrFe$_6$Ga$_6$ and type 1 LiMn$_6$Ge$_6$, and the corresponding magnetic moments calculated with the GGA and LDA functionals. Magnetic moments are given in $\mu_B$ per transition metal atom. A positive $J_n$ indicates a stable magnetic configuration, while a negative value indicates an unstable configuration. The corresponding Heisenberg model parameters are obtained by multiplying $J_n$ by minus for antiparallel pair alignment and dividing by $S_iS_j = M_iM_j/4$.

| System | Order | J$_1$/ meV | J$_2$/ meV | J$_3$/ meV | Moment/$\mu_B$ | |
|---|---|---|---|---|---|---|
| | | | | | GGA | LDA |
| ZrFe$_6$Ga$_6$ | FM | -10 | -10 | 117 | 1.91 | 1.75 |
| ZrFe$_6$Ga$_6$ | AFM1 | 20 | 20 | -42 | 1.92 | 1.74 |
| ZrFe$_6$Ga$_6$ | AFM2 | -43 | -33 | 58 | 1.91 | 1.76 |
| LiMn$_6$Ge$_6$ | FM | 16 | 32 | 8 | 2.21 | 2.00 |

### C. Microscopic exchange and breakdown of the Heisenberg picture

Our HTP survey reveals a broad distribution of energy differences between collinear magnetic states $\Delta = |\Delta E|$ across the AT$_6$X$_6$ family. To understand what controls the magnitude of $\Delta$ and how it correlates with the underlying magnetic ground state, we proceed from two complementary viewpoints. First, we analyze representative compounds using microscopic exchange parameters to determine whether a bilinear Heisenberg picture is adequate. Second, we relate these microscopic insights to the schematic spin-rotation energy landscapes in Fig. 3(c–e) and establish a practical $\Delta$-based heuristic for classifying magnetic stability across the full dataset. The magnetic phase diagram of FM kagome AT$_6$X$_6$ systems within a bilinear Heisenberg model has been discussed in Ref. [57] and is often used as a conceptual guide for the analysis of magnetic properties. However, its actual applicability for these AT$_6$X$_6$ systems has never been studied. This motivates us to examine the microscopic exchange interactions and evaluate how well a Heisenberg description captures the magnetic energetics of these materials.

We begin by examining the synthesized but magnetically uncharacterized type 4 ZrFe$_6$Ga$_6$ as a representative system with a small magnetic stability energy ($\Delta$ = 4.4 meV). At first glance, its magnetic behavior appears deceptively simple: the Fe moments vary by only about 10% in LDA and remain essentially unchanged across FM, AFM1, and AFM2 configurations, as shown in Table 2. Such behavior typically suggests localized moments and is often taken as evidence that a Heisenberg description should be applicable. However, the exchange parameters tell a very different story. The calculated RKKY-derived pairwise stability parameters [30,31] do not consistently support the presumably stable order and frequently indicate unstable spin alignments in nominal ground states. As shown in Fig. 4 and Table 2, the AFM2 configuration predicted by GGA, for example, exhibits negative $J_1$ and $J_2$, making it intrinsically incompatible with its assumed collinear structure. More broadly, in every collinear pattern (FM, AFM1, AFM2), at least one dominant interlayer coupling is negative. No collinear arrangement can satisfy all leading interactions simultaneously. The FM configuration reveals an even more striking anomaly: the next-nearest-neighbor interlayer coupling $J_3$ is large, positive, and exceeds $J_1$ and $J_2$ by more than an order of magnitude—directly opposing the typical RKKY decay with distance. This strongly distorted hierarchy signals that a short-ranged bilinear Heisenberg model is unlikely to be applicable in these systems, and more elaborate models including long-ranged, higher-order, and/or non-Heisenberg interactions are required.

In terms of magnetic energetics, this behavior maps naturally onto the landscape shown in Fig. 3(d). Here, strong frustration pushes competing FM and AFM states close in energy—an established hallmark of exchange competition and geometric frustration [58,59]. The resulting shallow spin-rotation landscape allows a nearby NC state to emerge as the actual ground state, making the collinear configurations appear nearly degenerate. As a contrasting example, we examine the newly predicted stable phase type 1 $LiMn_6Ge_6$, which emerges from our survey as a robust collinear ferromagnet. Here global screening yields a relatively large magnetic stability energy ($\Delta$ = 26.1 meV), and the stability parameters obtained are uniformly positive and remain stable with respect to the GGA-predicted ground state (Table 2). All leading couplings favor the same FM alignment, and no dominant interaction is frustrated. This behavior is fully consistent with a conventional Heisenberg description in which the FM configuration represents a deep, well-isolated energy minimum, as schematically depicted in Fig. 3(c).

Taken together, $ZrFe_6Ga_6$ and $LiMn_6Ge_6$ exemplify two qualitatively distinct microscopic regimes. The former displays pronounced non-Heisenberg, frustration-dominated behavior: several leading interlayer coupling parameters $J_n$ take on negative values that conflict with the imposed spin alignment across all collinear configurations. As a result, no collinear arrangement can simultaneously satisfy the stability of all dominant interactions, indicating a tendency toward a NC ground state (Fig. 3(d)). The latter exhibits conventional Heisenberg-like ferromagnetism, with all major $J_n$ cooperatively stabilizing the FM configuration and giving rise to a single, well-defined energetic minimum.

### D. Δ-based heuristic for magnetic stability across the $AT_6X_6$ family

Guided by the exchange-coupling analysis above, we now interpret the magnitude of $\Delta$ in terms of the schematic adiabatic spin-rotation energy landscapes shown in Fig. 5(a–c). In a conventional collinear Heisenberg magnet (Fig. 5(a)), the favored collinear state—FM or AFM—forms a single deep minimum, and the energy rises steeply as spins are rotated away from this orientation. $\Delta$ in this Heisenberg-like case is related to the critical temperature. The numerous experiments in the $AT_6X_6$ family clearly established the existence of rather high Curie or Néel temperatures, often well above room temperature, corresponding to characteristic energy scales on the order of 20–30 meV or higher. The case of an NC magnetic system is illustrated in Fig. 5(b). Here, neither FM nor AFM states are globally stable, and their stability parameters are negative with the energy minimum occurring at an intermediate rotation angle corresponding to some spiral or helical order. This situation is likely realized microscopically in $ZrFe_6Ga_6$, where the negative $J_n$ values preclude any collinear state from satisfying all dominant exchange interactions. In such cases, the $\Delta$ can be small or large, while the energy between the NC ground state and the closest collinear state is related to the observed temperature of magnetic transition of spin spiral to collinear state. Experimentally, such temperatures are often around 100-150 K.

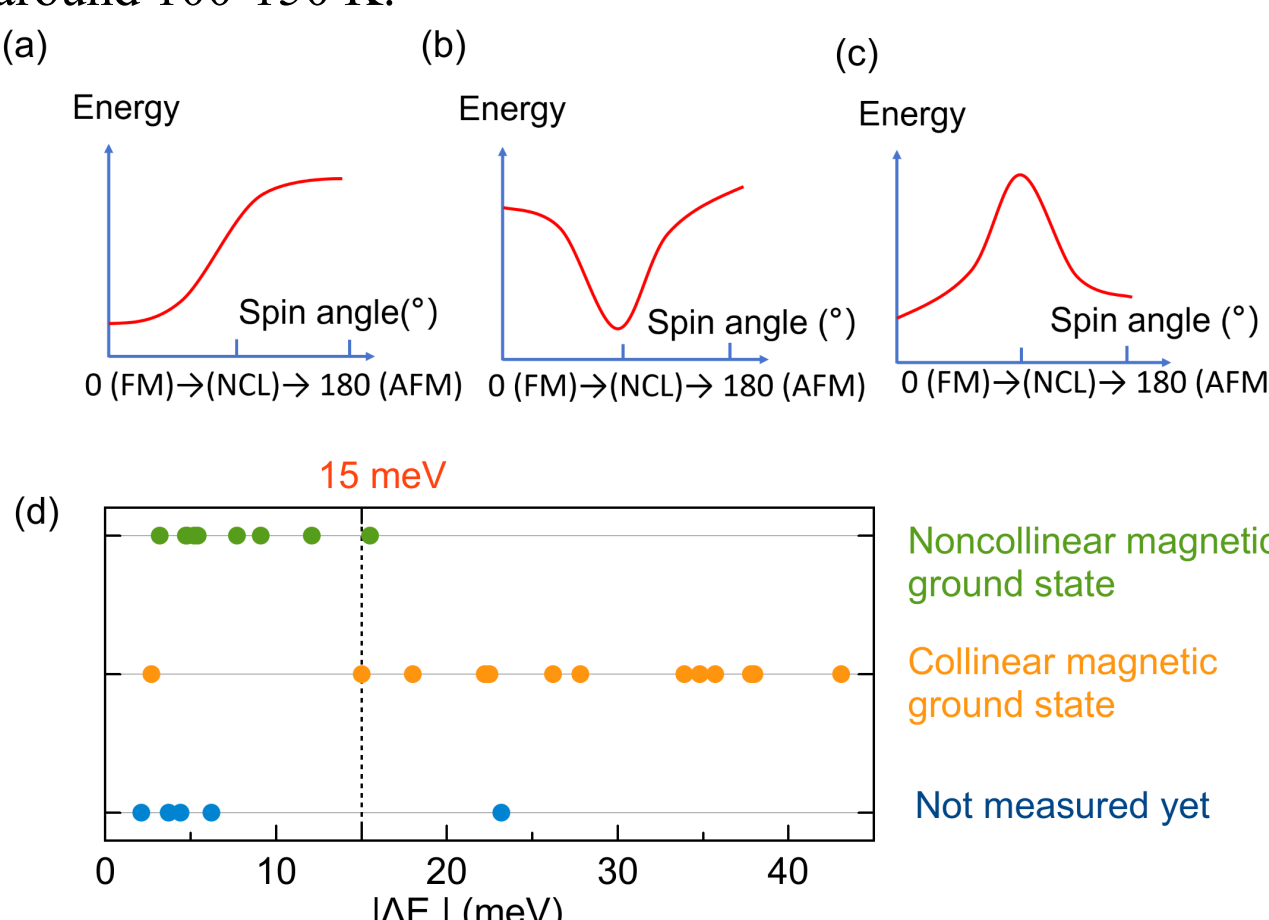

**FIG. 5** (a-c) Schematic spin-rotation energy landscapes illustrating different magnetic regimes. (a) Rigid collinear Heisenberg case: a single deep FM minimum and a steep energy rise with an angle, corresponding to large $|\Delta E|$. (b) Non-collinear regime: competing Heisenberg and/or non-Heisenberg interactions produce a non-collinear global minimum. (c) Nearly degenerate collinear states: both the FM (0°) and AFM (180°) configurations form distinct minima, separated by an energy barrier. (d) Correlation between the experimental magnetic ground state and the calculated energy difference $\Delta E$ for synthesized $AT_6X_6$ compounds. Circles represent individual systems, color-coded by their experimentally determined order: NC (green), collinear (orange), or undetermined (blue). The vertical dashed line at 15 meV marks the proposed stability threshold.

A third, intermediate scenario is depicted in Fig. 5(c), in which the FM and AFM states are separated by a modest NC energy barrier, resulting in possible nearly degenerate collinear states. However, to our knowledge, there are no experimental reports of such barriers in these systems. In contrast, the situation corresponding to energy surface on Fig.5(b) (appearance of stable NC configuration) has been observed many times in these systems. Thus, we assume that such near degeneracy most likely corresponds to traditional NC scenario of Fig.5(b).

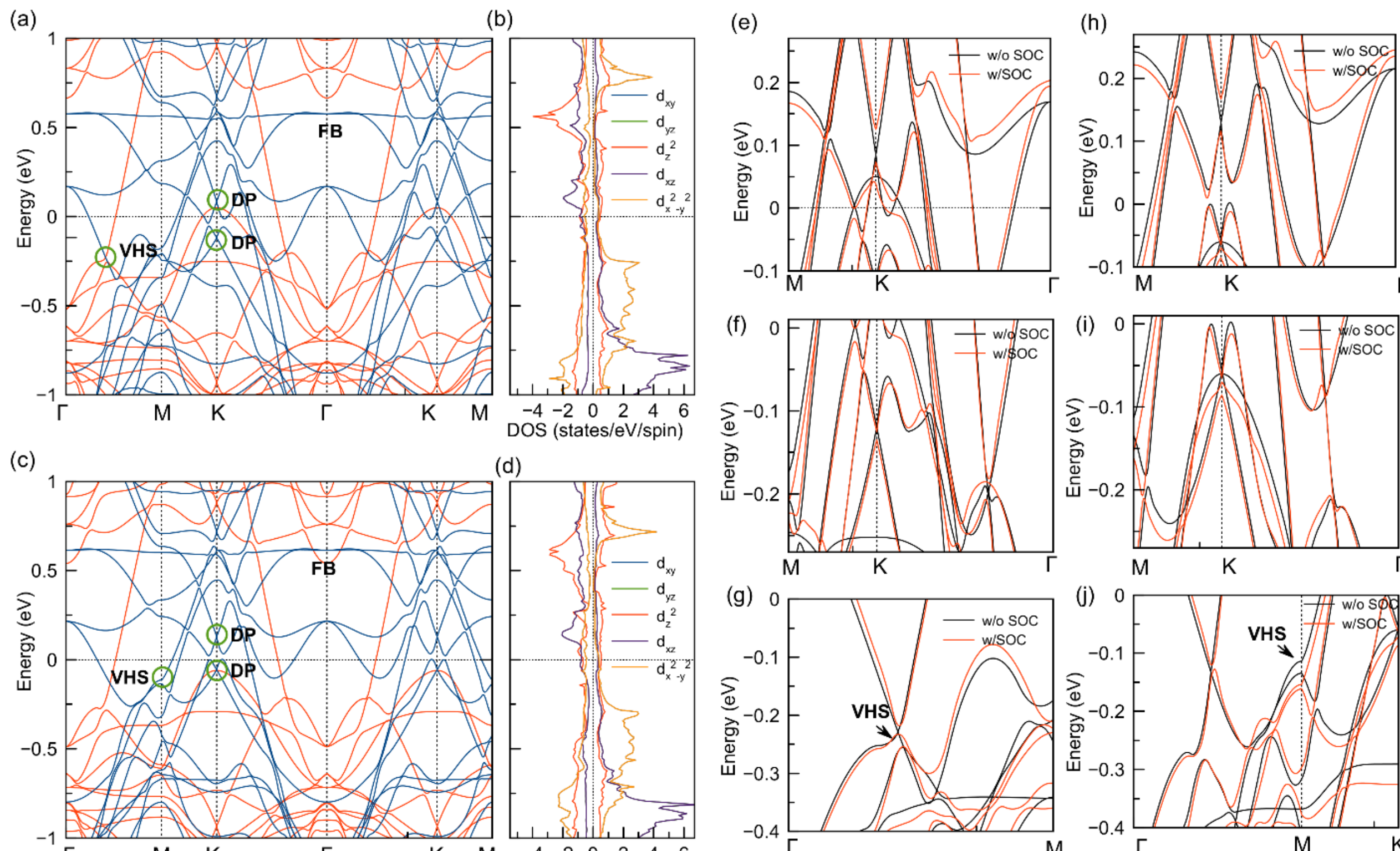

**FIG. 6** Spin-polarized band structures of (a) $LiMn_6Ge_6$ (type 1) and (c) $NaMn_6Ge_6$ (type 1) calculated without SOC. Red (blue) denote majority(minority) spins. Characteristic band features are highlighted by green circles: (a), Dirac-like crossings (DP) near K, and a FB along Γ–K–Γ. (b) The DOS of different d-orbitals of the Mn atoms in $LiMn_6Ge_6$. (d) The DOS of different d-orbitals of the Mn atoms in $NaMn_6Ge_6$. (e-j) compare band dispersions with and without SOC (black: w/o SOC; orange: w/ SOC) in a narrow energy window along M-K–Γ and Γ- M in (e-g) $LiMn_6Ge_6$ and (h-j) $NaMn_6Ge_6$.

Taking these considerations together, a small $\Delta$ signals that the FM and AFM states lie within a broader manifold of competing magnetic configurations. This insight is visually corroborated by Fig. 5(d), where the value $\Delta_c \approx 15$ meV emerges as a critical crossover value. We observe that all experimentally confirmed NC ground states (green circles) are confined below this threshold, while the region near 15 meV features both NC and collinear points, indicative of a phase boundary where competing interactions are nearly balanced (phase transition). Motivated by this distribution, we adopt 15 meV as a practical quantitative heuristic: systems with $\Delta$ clearly exceeding this threshold are classified as robust collinear magnets (populated exclusively by stable orange circles); whereas systems falling below or hovering near this limit are classified as frustration-prone. Notably, the experimentally FM $MgMn_6Sn_6$ falls into the near-degenerate regime with $\Delta = 2.7$ meV. This result underscores the utility of our classification scheme: although $MgMn_6Sn_6$ settles into an FM ground state, its small $\Delta$ accurately reflects the presence of strong magnetic frustration and competing exchange interactions. This distinguishes it from robust magnets like $LiMn_6Ge_6$ and aligns with recent theoretical studies [60], which independently characterize $MgMn_6Sn_6$ as a frustrated system with competing exchange.

Applying this criterion to the newly predicted stable compounds (Table 1) shows that $NaFe_6Ge_6$, $LiMn_6Ge_6$, and $NaMn_6Ge_6$ fall into the robust collinear regime, whereas $NbMn_6Ge_6$, $TiMn_6Ge_6$, $LaMn_6Ge_6$, and all $AFe_6Ga_6$ systems belong to the near-degenerate, frustration-prone class. This $\Delta$-based classification, anchored by the microscopic $J_n$ and the total energy analysis, provides a unified framework for interpreting the complex magnetic behavior of the $AT_6X_6$ family. Owing to its FM ground state and large $\Delta$, $LiMn_6Ge_6$—together with $NaMn_6Ge_6$, which our calculations likewise identify as a stable FM system—emerges as an ideal platform for exploring the topological electronic structure of kagome ferromagnets. To further assess the feasibility of these two compounds, we carried out phonon and elastic-constant calculations for $LiMn_6Ge_6$ and $NaMn_6Ge_6$. The absence of imaginary phonon branches and the satisfaction of the Born stability criteria confirm that both systems are dynamically and mechanically stable within the present DFT framework (see Table S8 and Fig. S5). In the following, we present a detailed electronic-structure analysis of these two compounds.

### E. Electronic structure analysis

Given that robust ferromagnetism is of primary interest. Below we focus our electronic–structure analysis on two type 1 kagome metals with robust collinear FM ground states, $LiMn_6Ge_6$ and $NaMn_6Ge_6$. Their spin-polarized band structures and Mn-d projected DOS are shown in Fig. 6. We first examine the electronic structure of stable FM phase $LiMn_6Ge_6$ in Fig. 6(a). The bands display a spin-polarized kagome motif. Without SOC, two minority-spin Dirac-like crossings are resolved near K at +0.085 eV and −0.126 eV. A pronounced majority-spin van Hove singularity occurs around −0.24 eV at the midpoint of Γ–M, dominated by Mn $d_{x^2-y^2}$ character, consistent with the PDOS in Fig. 6(b). In addition, a kagome-derived flat band appears along Γ–K–M–Γ at approximately 0.55 eV, mainly originating from Mn $d_{z^2}$ states. Including SOC slightly shifts this flat band downward to ≈0.53 eV (Fig. S3) and modifies the other key features: the two crossings acquire gaps of 52 meV and 10 meV (Figs. 6e and 6f), and the saddle-point degeneracy is lifted by a small, avoided crossing of 14 meV (Fig. 6g). Notably, the 52 meV SOC gap is sizable among kagome magnets; it is on par with theoretical estimates of the Chern gap in $TbMn_6Sn_6$ (~55 meV) [61], indicating that $LiMn_6Ge_6$ is a strong candidate for realizing quantized or strongly enhanced anomalous Hall and Nernst responses when the chemical potential is tuned near the gapped Dirac cones or the VHS.

$NaMn_6Ge_6$ shows a closely related yet distinct kagome electronic structure [Figs. 6(c,d)]. In this compound all characteristic kagome features originate from the minority-spin channel: both Dirac-like crossings near K, the VHS near M, and the FB along Γ–K–M–Γ are carried by minority-spin bands, whereas the majority-spin states around $E_F$ are more dispersive. Without SOC, the two Dirac points sit very close to the Fermi level at approximately +0.13 and −0.06 eV. A sharp VHS appears at the M point only 0.011 eV below $E_F$, and the FB lies at about +0.60 eV. Thus, in $NaMn_6Ge_6$ the VHS is almost pinned to the Fermi level, while the Dirac crossings and FB remain in its immediate vicinity. SOC has a similar qualitative effect in $NaMn_6Ge_6$ [Figs. 6(h–j)]. The two Dirac-like crossings acquire finite gaps of roughly 51 and 8 meV, respectively. At the same time the VHS at M is pushed downward by about 40 meV, but it still resides close to the Fermi level. Because the Dirac points, VHS, and FB are all hosted by the same minority-spin manifold, SOC is expected to generate substantial and predominantly same-sign Berry curvature concentrated in this spin channel, which is favorable for producing sizable anomalous Hall and thermoelectric responses upon moderate carrier doping.

Thus, both $LiMn_6Ge_6$ and $NaMn_6Ge_6$ realize a "Dirac–VHS–flat-band" landscape characteristic of kagome metals and emerge as prime candidates for experimental exploration of large anomalous Hall and Nernst effects and for engineering topological transport via modest tuning of the chemical potential.

## IV. Conclusions

In summary, our high-throughput studies of the $AT_6X_6$ family unveil a complex magnetic landscape. Beyond identifying nine new stable structural phases (to the best of our knowledge), we predict magnetic ground states for a broad range of metastable candidates. Among them, we predict $LiMn_6Ge_6$ and $NaMn_6Ge_6$ to be stable ferromagnets and $NaFe_6Ge_6$ to be a stable antiferromagnet, while six other candidates exhibit much lower magnetic stability regarding their collinear states. We propose a simple criterion, supported by total energy analysis, to distinguish robust collinear orders from potentially frustrated cases. Our studies confirm that the stable ferromagnets, $LiMn_6Ge_6$ and $NaMn_6Ge_6$, exhibit gapped Dirac cones near $E_F$ and a nearby VHS, which can be accessed by modest carrier doping, establishing them as promising candidates for realizing large anomalous Hall and Nernst effects. Overall, our work provides a predictive map of crystalline and magnetic stabilities for the $AT_6X_6$ kagome family, offering distinct pathways for exploring both collinear topological magnetism and non-collinear spintronics. Future research beyond the present collinear DFT framework, incorporating non-collinear magnetism, finite-temperature effects, and electronic correlations will be an important direction for future work.

**Acknowledgement**

The work at Xiamen University was supported by the National Natural Science Foundation of China (Grant No. T2422016), the Natural Science Foundation of Xiamen (Grant No. 3502Z202371007), and the Fundamental Research Funds for the Central Universities (Grant No. 20720230014). The work of V.A. and Z.Z. was supported by the DOE Established Program to Stimulate Competitive Research (EPSCoR) Grant No. DE-SC0024284. Ames National Laboratory is operated for the U.S. Department of Energy by Iowa State University under Contract No. DE-AC02-07CH11358.

**Data availability**

The data that support the findings of this article are openly available [62].